\documentclass[conference]{IEEEtran}
\IEEEoverridecommandlockouts

\newfont{\bb}{msbm10 scaled 1100}

\def\tr{\mathrm{tr}}

\def\diag{\mathrm{diag}}
\usepackage{graphicx}
\usepackage{algorithm,algorithmic}
\usepackage{psfrag}
\usepackage{url}
\usepackage{amsmath}
\usepackage{amsfonts,amssymb,amsmath}
\usepackage{mathrsfs}
\usepackage{amsthm}
\usepackage{array}
\usepackage{cite}
\usepackage{color}

\newtheorem{theorem}{Theorem}

\newtheorem{lemma}{Lemma}

\newtheorem{corollary}{Corollary}

\newcommand{\vect}[1]{\mathbf{#1}}

\newcommand{\minimize}[1]{{\underset{{#1}}{\mathrm{minimize}}}}

\begin{document}
\IEEEoverridecommandlockouts
%
\title{{Optimal Linear Precoding in Multi-User MIMO Systems: A Large System Analysis}\vspace{-0.2cm}}
\author{
\IEEEauthorblockN{Luca~Sanguinetti\IEEEauthorrefmark{1}\IEEEauthorrefmark{3},
Emil~Bj{\"o}rnson\IEEEauthorrefmark{3}\IEEEauthorrefmark{2}, M{\'e}rouane Debbah\IEEEauthorrefmark{3} and  Aris~L. Moustakas\IEEEauthorrefmark{3}\IEEEauthorrefmark{4}
\thanks{L.~Sanguinetti is funded by the People Programme (Marie Curie Actions) FP7 PIEF-GA-2012-330731 Dense4Green. E.~Bj\"ornson is funded by an International Postdoc Grant from the Swedish Research Council. A. L. Moustakas is the holder of the DIGITEO ``ASAPGONE'' Chair. This research has also been supported by
the FP7 NEWCOM\# (Grant no. 318306), the ERC Starting MORE (Grant no.
305123), the French p\^ole de comp\'etitivit\'e SYSTEM@TIC within the
project 4G in Vitro.}
}
\IEEEauthorblockA{\IEEEauthorrefmark{1}\small{Dipartimento di Ingegneria dell'Informazione, University of Pisa, Pisa, Italy}}
\IEEEauthorblockA{\IEEEauthorrefmark{3}\small{Alcatel-Lucent Chair, Ecole sup{\'e}rieure d'{\'e}lectricit{\'e} (Sup{\'e}lec), Gif-sur-Yvette, France}}
\IEEEauthorblockA{\IEEEauthorrefmark{2}\small{Dept. of Signal Processing, KTH, Stockholm, and Dept. of Electrical Engineering, Link\"{o}ping University, Sweden}}
\IEEEauthorblockA{\IEEEauthorrefmark{4}\small{Department of Physics,
National \& Capodistrian University of Athens, Athens, Greece}}
\vspace{-.8cm}}
\maketitle

\begin{abstract}
We consider the downlink of a single-cell multi-user MIMO system in which the base station makes use of $N$ antennas to communicate with $K$ single-antenna user equipments (UEs) randomly positioned in the coverage area. In particular, we focus on the problem of designing the optimal linear precoding for minimizing the total power consumption while satisfying a set of target signal-to-interference-plus-noise ratios (SINRs). To gain insights into the structure of the optimal solution and reduce the computational complexity for its evaluation, we analyze the asymptotic regime where $N$ and $K$ grow large with a given ratio and make use of recent results from large system analysis to compute the asymptotic solution. Then, we concentrate on the asymptotically design of heuristic linear precoding techniques. Interestingly, it turns out that the regularized zero-forcing (RZF) precoder is equivalent to the optimal one when the ratio between the SINR requirement and the average channel attenuation is the same for all UEs. If this condition does not hold true but only the same SINR constraint is imposed for all UEs, then the RZF can be modified to still achieve optimality if statistical information of the UE positions is available at the BS. Numerical results are used to evaluate the performance gap in the finite system regime and to make comparisons among the precoding techniques.
\end{abstract}

\vspace{-0.2cm}
\section{Introduction}

Multiple-Input Multiple-Output (MIMO) technologies are currently being adopted in many wireless communication standards such as fourth generation (4G) cellular networks \cite{Qinghua2010}. The main limiting factor in multi-user MIMO systems is the multiple-access interference (MAI). In uplink transmissions, MAI mitigation is typically accomplished at the base station (BS) using linear multi-user detectors or non-linear techniques based on layered architectures.
In downlink transmissions, MAI mitigation can only be accomplished at the BS using precoding techniques. As shown in \cite{Weingarten2006}, the capacity-achieving precoding strategy is dirty paper coding (DPC). Although optimal, the implementation of DPC requires a tremendous computational complexity at both BS and user equipments (UEs). On the other hand, a practical approach that has received considerable attention (due to its simplicity) is represented by linear precoding or beamforming \cite{Gershman2010,Bjornson:2014mag}.

In this work, we focus on the problem of designing the optimal linear precoding for minimizing the total transmit power while satisfying a set of target signal-to-interference-plus-noise ratios (SINRs) \cite{Schubert2004a,Wiesel06,Yu:2007ja,Bjornson2012a}. This problem is receiving renewed interest nowadays due to the emerging research area of green cellular networks \cite{Chen2011}. In particular, we consider the downlink of a single-cell multi-user MIMO system in which the BS makes use of $N$ antennas to communicate with $K$ single-antenna UEs randomly positioned in the coverage area. Under the assumption of perfect channel state information (CSI), the solution to the power minimization problem in this context was originally computed in \cite{Bengtsson2001a} and later extended to different scenarios in \cite{Wiesel06,Yu:2007ja,Bjornson2012a}. In particular, it turns out that the optimal linear precoder depends on some Lagrange multipliers whose computation can be performed using convex optimization tools or solving a fixed-point problem \cite{Bjornson:2014mag}. Although possible, both approaches do not provide any insights into the structure of the optimal values. Moreover, the computation must be performed for any new realization of the MIMO channel matrix.

To overcome these issues, we follow the same approach as in \cite{Lakshminaryana2010,Huang2012,Zakhour2012,Asgharimoghaddam14} and resort to the asymptotic regime where $N$ and $K$ grow large with a given ratio $c = K/N$. Differently from \cite{Lakshminaryana2010,Huang2012,Zakhour2012,Asgharimoghaddam14}, the asymptotically optimal values of the Lagrange multipliers are computed using the approach adopted in \cite{Romain2014}, which provides us a much simpler means to overcome the technical difficulties arising with the application of standard random matrix theory tools (see for example \cite{Zakhour2012}). As already pointed out in \cite{Lakshminaryana2010,Huang2012,Zakhour2012,Asgharimoghaddam14}, in the asymptotic regime the optimal values can be computed in closed-form through a nice and simple expression, which depends only on the user positions and SINR requirements.
The above results are then used to validate the optimality of different heuristic linear precoding techniques, which are inspired by the widely used regularized zero-forcing (RZF) concept \cite{Wagner12,Evans2013,Bjornson:2013kc} and its extensions to include arbitrary user priorities \cite{Bjornson:2014mag}.
The optimal regularization parameter is provided in the asymptotic regime. To the best of authors' knowledge, this is the first time that such a result is found since most of the related works are focused on sum rate maximization. Comparisons are then made with two heuristic techniques. The former is the classical RZF precoder \cite{Wagner12} while the latter is referred to as position-aware RZF (PA-RZF) precoder since it relies on knowledge of the UE positions \cite{Evans2013}. Interestingly, it turns out that PA-RZF is equivalent to the optimal linear precoder when the same SINR constraint is imposed for all UEs. On the other hand, the commonly used RZF precoder becomes optimal only when the ratio between the SINR requirement and the average channel attenuation is the same for all UEs. Numerical results are used to evaluate the performance gap in the finite system regime and to make comparisons among the different precoding techniques.

\section{System Model and Problem Formulation}



We consider the downlink of a single-cell multi-user MIMO system in which the BS makes use of $N$ antennas to communicate with $K$ single-antenna UEs. The $K$ active UEs change over time and are randomly selected from a large set of UEs within the coverage area. The physical location of UE $k$ is denoted by $\mathbf{x}_k \in \mathbb{R}^2$ (in meters) and it is computed with respect to the BS (assumed to be located in the origin).
The function $l(\cdot)\!\!\!: \mathbb{R}^2 \rightarrow \mathbb{R}_+$ describes the large-scale channel fading at different user locations; that is, $l(\mathbf{x}_k )$ is the average channel attenuation due to path-loss and shadowing at location $\mathbf{x}_k$. The large-scale fading between a UE and the BS is assumed to be the same for all BS antennas. This is reasonable since the distances between UEs and BS are much larger than the distance between the BS antennas. Since the forthcoming analysis does not depend on a particular choice of $l(\cdot)$, we keep it generic. Perfect CSI is assumed to be available at the BS for analytic tractability. The imperfect CSI case is left for future work.\footnote{We limit to observe that might in principle be included following the same approach adopted in \cite{Wagner12}. See also \cite{Sanguinetti_JSAC_14}.}


The BS shall convey the information symbol $s_k$ to UE $k$ using linear precoding. The symbol vector $\mathbf{s} = [s_1,s_2,\ldots,s_K]^T \in \mathbb{C}^{K\times 1}$ originates from a Gaussian codebook with zero mean and covariance matrix $\mathbb{E}_{\mathbf{s}}[\mathbf{s}\mathbf{s}^H] = \mathbf{I}_K$. Denoting by $\mathbf{V} = [\mathbf{v}_1,\mathbf{v}_2,\ldots,\mathbf{v}_K] \in \mathbb{C}^{N\times K}$ the precoding matrix, the received sample ${y}_{k} \in \mathbb{C}$ at UE $k$ takes the form
\begin{align}\label{y_k}
{y}_{k}=\mathbf{h}_{k}^H\mathbf{V}\mathbf{s}+{n}_{k}
\end{align}
where ${n}_{k}\sim \mathcal {CN} (0,\sigma^2)$ is the additive noise and the entry $h_{k,n}$ of the row vector $\mathbf{ h}_{k}^H= [{ h}_{k,1},{ h}_{k,2},\ldots,{ h}_{k,N}] \in \mathbb{C}^{1 \times N}$ is the channel propagation coefficient between the $n$th antenna at the BS and the $k$th UE. We assume a Rayleigh fading channel model ${\mathbf{h}}_{k}= \sqrt{l(\mathbf{x}_{k})} \mathbf{w}_k$
with ${\mathbf{w}}_k\sim \mathcal {CN} (0,\mathbf{I}_N)$ accounting for the  small-scale fading channel. The SINR at the $k$th UE is easily written as \cite{Bjornson:2013kc}
\begin{equation}\label{1}
{\rm{SINR}}_k = \frac{\left|\mathbf{h}_k^H\mathbf{v}_k\right|^2}{\sum\limits_{i=1,i\ne k}^K\left|\mathbf{h}_k^H\mathbf{v}_i\right|^2 + {\sigma^2}}.
\end{equation}
As mentioned earlier, we consider the power minimization problem whose mathematical formulation is as follows:
\begin{align}\label{Opt_problem}
\mathcal P: \quad\quad \minimize{\mathbf{V}}& \quad \quad P = \tr ( \mathbf{V}\mathbf{V}^H ) \\ \label{SINR_constraints}
{\text{subject to}} & \quad \quad {\rm{SINR}}_k \ge \gamma_k \quad k=1,2,\ldots,K
\end{align}
where $\gamma_k$ is the given SINR target of UE $k$ obtained as (under the assumption of Gaussian codebooks) $\gamma_k = 2^{r_k}-1$
with $r_k$ being the target user rate in bit/s/Hz. For later convenience, we call $\boldsymbol{\gamma} = [\gamma_1,\gamma_2,\ldots,\gamma_K]^T$.

\section{Optimal Linear Precoding}\label{Section_optimal}

As originally shown in \cite{Bengtsson2001a}, the non-convex optimization problem $\mathcal P$ in \eqref{Opt_problem} can be put in a convex form by reformulating the SINR constraints as second-order cone constraints. In doing so, the optimal $\mathbf{V}^\star$ is found to be \cite{Wiesel06,Yu:2007ja,Bjornson2012a}
\begin{align}\label{22}
\mathbf{V}^\star=\left(\sum\limits_{i=1}^K\lambda_i^\star\mathbf{  h}_i\mathbf{  h}_i^H + N\mathbf{I}_N\right)^{-1}\mathbf{  H} \sqrt{\mathbf{P}^{\star}}
\end{align}
where $\mathbf{H}=[\mathbf{  h}_1,\mathbf{  h}_2,\ldots,\mathbf{  h}_K] \in \mathbb{C}^{N\times K}$ and $\boldsymbol{\lambda}^\star= [\lambda_1^\star,\lambda_2^\star,\ldots,\lambda_K^\star]^T$ is the positive unique fixed point of the following equations \cite{Schubert2004a,Wiesel06,Bjornson:2014mag}:
\begin{align}\label{23}
\left(1 + \frac{1}{\gamma_k}\right) \lambda_k^\star= \frac{1}{\mathbf{h}_k^H\left(\sum\limits_{i=1}^K\lambda_i^\star\mathbf{  h}_i\mathbf{  h}_i^H + N\mathbf{I}_N\right)^{-1}\mathbf{h}_k}
\end{align}
for $k=1,2,\ldots,K$. Also, $\mathbf{P}^{\star}=\diag\{p_1^\star,p_2^\star,\ldots,p_K^\star\}$ is a diagonal matrix whose entries are such that the SINR constraints in $\mathcal P$ are all satisfied with equality when $\mathbf{V} = \mathbf{V}^\star$. Plugging \eqref{1} into \eqref{SINR_constraints}, the optimal vector $\mathbf{p}^\star = [p_1^\star,p_2^\star,\ldots,p_K^\star]^T$ is computed as \cite{Bjornson:2014mag}
\begin{align}\label{24}
\mathbf{p}^\star = \sigma^2 \mathbf{D}^{-1} \mathbf{1}_{K}
\end{align}
where the $(k,i)$th element of $\mathbf{D} \in \mathbb C^{K\times K}$ is
\begin{align}\label{25}
\left[\mathbf{D}\right]_{k,i}= \begin{cases}
\frac{1}{\gamma_k}{|\vect{h}_k^H \vect{a}_k^\star|^2}& \text{for} \,\,\, k = i \\
-{|\vect{h}_k^H \vect{a}_{i}^\star|^2}& \text{for} \,\,\, k \ne i
\end{cases}
\end{align}
with $ \vect{a}_k^\star$ being the $k$th column of $\vect{A}^\star = (\sum\nolimits_{i=1}^K\lambda_i^\star\mathbf{  h}_i\mathbf{  h}_i^H + N\mathbf{I}_N)^{-1}\mathbf{  H}$.

As seen, $\mathbf{V}^\star$ in \eqref{22} is parameterized by $\boldsymbol{\lambda}^\star$ and $\mathbf{p}^\star$, where $\boldsymbol{\lambda}^\star$ needs to be evaluated by an iterative procedure due to the fixed-point equations in \eqref{23}. This is a computationally demanding task when $N$ and $K$ are large since the matrix inversion operation in \eqref{23} must be recomputed at every iteration and its computational complexity scales proportionally to $N^2 K$. Moreover, computing $\boldsymbol{\lambda}^\star$ as the fixed point of \eqref{23} does not provide any insights into the optimal structure of both $\boldsymbol{\lambda}^\star$ and $\mathbf{p}^\star$.  In addition, the parameter values depend directly on the channel vectors $\{\mathbf{h}_i\}$ and change at the same pace as the small-scale fading (i.e., at the order of milliseconds).

To overcome the above issues, we assume that $N,K \rightarrow \infty$ with $K/N=c \in (0,1]$ and use some recent tools in large system analysis to compute the so-called deterministic equivalents of $\boldsymbol{\lambda}^\star$ and $\mathbf{p}^\star$. For later convenience, we call
\begin{equation}\label{26}
\xi = 1 - \frac{1}{N}\sum\limits_{i=1}^K \frac{\gamma_i}{1+\gamma_i} \quad \text{and} \quad
A =  \frac{1}{K} \sum\limits_{i=1}^K \frac{\gamma_i}{ l(\mathbf{x}_{i})}.
\end{equation}

\subsection{Asymptotically Optimal Linear Precoding}

The following theorem provides the solution to the optimization problem in \eqref{Opt_problem} in the asymptotic regime.

\begin{theorem} \label{lemma:asymptotic-beamforming}
If $N,K \rightarrow \infty$ with $c \in (0,1]$, then
\begin{equation}\label{27}
\mathop {\max }\limits_{k = 1,2,\ldots,K} \left| \lambda_k^\star  - \overline \lambda_k\right| \mathop {\longrightarrow}\limits^{a.s.} 0
\end{equation}
and
\begin{equation}\label{28}
\mathop {\max }\limits_{k = 1,2,\ldots,K} \left|p_k^\star  - \overline p_k \right|\mathop {\longrightarrow}\limits^{a.s.} 0
\end{equation}
where $\overline \lambda_k$ and $ \overline p_k$ are the deterministic equivalents of $\lambda_k^\star$ and $ p_k^\star$, respectively, and are given by
\begin{equation}\label{29}
\overline \lambda_k =  \frac{\gamma_k}{l(\mathbf{x}_{k})\xi}
\end{equation}
and
\begin{equation}\label{30}
\overline p_k = \frac{\gamma_k}{l(\mathbf{x}_{k})\xi^2}\left(\overline P + \frac{\sigma^2}{l(\mathbf{x}_{k})} \left(1+\gamma_k\right)^2\right)
\end{equation}
with
\begin{equation}\label{31}
\overline P = \frac{cA\sigma^2}{\xi}
\end{equation}
being the deterministic equivalent of the transmit power $P$.
\end{theorem}
\begin{IEEEproof}
Similar results have previously been derived by applying standard random matrix theory tools to the right-hand-side of \eqref{23}. However, the application of these tools to the problem at hand is not analytically correct since the Lagrange multipliers in \eqref{23} are a function of the channel vectors $\{{\bf{h}}_k\}$. To overcome this issue, we make use of the same approach adopted in \cite{Romain2014} whose main steps are sketched in \cite{Sanguinetti_2014_Mobility}. On the other hand,  \eqref{30} is proved using standard random matrix theory results (omitted for space limitations).
\end{IEEEproof}

The following remarks elaborate on some of the insights that are obtained from Theorem \ref{lemma:asymptotic-beamforming}.

{\bf{Remark 1}}. In sharp contrast to \eqref{23}, the computation of $\overline \lambda_k$ in \eqref{29} only requires knowledge of the user position through $l(\mathbf{x}_{k})$. This information can be easily observed and estimated accurately at the BS because it changes slowly with time (relative to the small-scale fading). The Lagrange multiplier $\lambda_k$ is known to act as a user priority parameter that implicitly determines how much interference the other UEs may cause to UE $k$ \cite{Bjornson:2014mag}. Interestingly, its asymptotic value $\overline \lambda_k$ is proportional to the SINR $\gamma_k$ and inversely proportional to $l(\mathbf{x}_{k})$ such that users with weak channels have larger values.
Higher priority is thus given to users that require high performance and/or have weak propagation conditions \cite{Bjornson:2014mag}.

{\bf{Remark 2.}} A known problem with using the asymptotically optimal power allocation in Lemma \ref{lemma:asymptotic-beamforming} is that the target SINRs are not guaranteed to be achieved at finite numbers of antennas (see for example \cite{Asgharimoghaddam14}). This is because the approximation errors are translated into fluctuations in the resulting SINR values. However, these errors rapidly vanish also in the finite regime when $N$ is larger than $K$, which is the regime envisioned for massive MIMO systems \cite{Rusek2013a}. It can also be avoided by using only the deterministic equivalents $\overline \lambda_k $ of the Lagrange multipliers and computing the power allocation coefficients according to \eqref{24}. This approach retains most of the complexity benefits of the asymptotic analysis.

The following corollaries can be easily obtained from Theorem \ref{lemma:asymptotic-beamforming} and will be useful later on.
\begin{corollary}\label{corollary_OPL_same_ratios}
If the ratio between the SINR requirement and the average channel attenuation is the same for all UEs and equal to some $\zeta \ge 0$, i.e.,
\begin{align}\label{zeta}
\frac{\gamma_k}{l(\mathbf{x}_{k})} =\zeta \quad \quad k = 1,2,\ldots,K,
\end{align}
then $\overline \lambda_k$ in \eqref{29} takes the form
\begin{equation}\label{26.2.D}
\overline \lambda_k =  \frac{ \zeta}{\xi} = \zeta \left( 1 - \frac{1}{N}\sum\limits_{i=1}^K \frac{\gamma_i}{1+\gamma_i}\right)^{-1}.
\end{equation}
\end{corollary}
\begin{corollary}[\!\!\cite{Zakhour2012}]\label{corollary_OPL_same_rates}
If the same target SINR is imposed for each user, i.e.,
\begin{equation}\label{gamma}
\boldsymbol{\gamma} = \gamma \boldsymbol{1}_k,
\end{equation}
then $\overline \lambda_k$ in \eqref{29} reduces to
\begin{equation}\label{26.2}
\overline \lambda_k = \frac{\gamma}{l(\mathbf{x}_{k})\xi }= \frac{\gamma}{l(\mathbf{x}_{k})}\left( 1 - c \frac{\gamma}{1+\gamma}\right)^{-1}
\end{equation}
and $\overline P$ becomes
\begin{equation}\label{26.3}
\overline P = {cA\sigma^2}\left( 1 - c\frac{\gamma}{1+\gamma}\right)^{-1}.
\end{equation}
\end{corollary}



\section{Heuristic Linear Precoding}

Inspired by the optimal linear precoding in \eqref{22}, we now consider suboptimal precoding techniques that builds on heuristics \cite{Bjornson:2014mag}.
To this end, we let $\mathbf{V}$ take the following general form
\begin{align}\label{22.101}
\mathbf{V}&=\left(\sum\limits_{i=1}^K\alpha_i\mathbf{  h}_i\mathbf{  h}_i^H + N\rho\mathbf{I}_N\right)^{-1}\mathbf{  H}\sqrt{\mathbf{  P}}
\end{align}
where $\boldsymbol{\alpha}= [\alpha_1,\alpha_2,\ldots,\alpha_K]^T$ is now a given vector with positive scalars and $\rho$ is a design parameter to be optimized. Note that \eqref{22.101} is basically obtained from \eqref{22} by setting $\lambda_k = {\alpha_k}/{\rho}$ for all $k$. As before, the power allocation $\mathbf{P}$ is computed according to \eqref{24} and satisfies all the SINR constraints with equality.

Observe that if $\boldsymbol{\alpha}$ is set to $\boldsymbol{1}_K$, then $\mathbf{V}$ in \eqref{22} reduces to the well-known RZF precoder \cite{Wagner12}:
\begin{align}\label{28}
\mathbf{V}_{\rm{RZF}}&=\left(\sum\limits_{i=1}^K\mathbf{  h}_i\mathbf{  h}_i^H + N\rho\mathbf{I}_N\right)^{-1}\mathbf{  H}\sqrt{\mathbf{  P}}.
\end{align}
This particular precoding matrix is also known as the transmit Wiener filter and signal-to-leakage-and-noise ratio (SLNR) maximizing beamforming (see Remark 3.2 in \cite{Bjornson:2013kc} for a historical exposition).

On the other hand, if the BS makes use of knowledge of the user positions and let $\boldsymbol{\alpha}$ be equal to $\mathbf{L}^{-1}\mathbf{1}_K$ with $\mathbf{L} = \diag\{l(\mathbf{x}_{1}),l(\mathbf{x}_{2}),\ldots,l(\mathbf{x}_{K})\}$, then the processing matrix $\mathbf{V}$ in \eqref{22} reduces to (see also \cite{Evans2013,Sanguinetti_2014_Mobility,Sanguinetti_JSAC_14})
\begin{align}\label{31}
\mathbf{V}_{\rm{PA-RZF}}&=\left(\sum\limits_{i=1}^K\mathbf{  w}_i\mathbf{  w}_i^H + N\rho\mathbf{I}_N\right)^{-1}\mathbf{  H}\sqrt{\mathbf{  P}}
\end{align}
which we refer to as PA-RZF precoder in the sequel.

Differently from the optimal linear precoding that requires to compute the fixed point of a set of equations, the optimization of a linear precoder in the form of \eqref{22.101} requires only to look for the value of $\rho$ minimizing the transmit power. This can generally not be done in closed-form but requires a numerical optimization procedure \cite{Bjornson:2014mag}. To overcome this problem, the asymptotic regime is analyzed in the sequel.

\subsection{Asymptotic Analysis of the Heuristic Linear Precoding}

We keep $\boldsymbol{\alpha}$ generic and look for the value of $\rho$ that minimizes the total transmit power $P$ in \eqref{Opt_problem} when $N,K \rightarrow \infty$ with $c \in (0,1]$. In doing so, the following result is obtained.

\begin{theorem}\label{theorem_general_V}
If $N,K \rightarrow \infty$ with $c \in (0,1]$, then the parameter $\rho$ minimizing the deterministic equivalent of $P$ with $\mathbf{V}$ given by \eqref{22} is
\begin{align}\label{23.1}
\rho^\star = \frac{1}{\mu^\star} - \frac{1}{N}\sum\limits_{i=1}^K\frac{\alpha_il(\mathbf{x}_{i})}{1+\alpha_il(\mathbf{x}_{i}) \mu^\star}
\end{align}
where $\mu^\star$ is the solution of the following fixed point equation:
\begin{align}\label{24.fixedpoint}
\!\!\!\mu^\star \!\!= \!\!\left(\sum\limits_{i=1}^K \frac{\alpha_il(\mathbf{x}_{i})\gamma_i}{\left(1 + \alpha_il(\mathbf{x}_{i})\mu^\star\right)^3}\right)\!\!\!\left(\sum\limits_{i=1}^K \frac{\left(\alpha_il(\mathbf{x}_{i})\right)^2}{\left(1 + \alpha_il(\mathbf{x}_{i})\mu^\star\right)^3}\right)^{-1}\!\!\!\!\!\!.
\end{align}
In addition, the deterministic equivalent of $p_k$ takes the form
\begin{align}
\overline p_k =  \frac{\gamma_k}{l(\mathbf{x}_{k})(\mu^{\star})^2}{{\left[\overline P + \frac{\sigma^2}{l(\mathbf{x}_{k})} \left(1+\alpha_kl(\mathbf{x}_{k})\mu^{\star}\right)^2\right]}}{}
\end{align}
where 
\begin{align}\label{36.100}
\overline P= \frac{cA\sigma^2}{1 - (\mu^\star)^2 F_2 - cB}
\end{align}
is the deterministic equivalent of transmit power with $A$ given by \eqref{26} and
\begin{align}\label{36.1001}
B = \frac{1}{K}\sum\limits_{i=1}^K\frac{\gamma_i}{\left(1 + \alpha_il(\mathbf{x}_{i})\mu^\star\right)^2}\\\label{36.1002}
F_2= \frac{1}{N}\sum\limits_{i=1}^{K}\frac{(\alpha_il(\mathbf{x}_{i}))^2}{\left(1+\alpha_il(\mathbf{x}_{i})\mu^\star\right)^2}.
\end{align}
\end{theorem}
\begin{IEEEproof}
The proof is detailed in Appendix and operates in two steps. In the first step, we use the results of Theorem 1 in \cite{Wagner12} to compute $\overline {\rm{SINR}}_k$ and $\overline P$, i.e., the deterministic equivalents of ${\rm{SINR}}_k$ and $P$, respectively. In the second step, we set ${\rm{\overline {SINR}}}_k= \gamma_k$ for $k=1,2,\ldots,K$ and compute the corresponding powers $\{\overline p_k\}$, which are eventually used to obtain $\overline P$. The latter takes the form in \eqref{36.100} from which taking the derivative with respect to $\rho$ we obtain \eqref{23.1} and \eqref{24.fixedpoint}.
\end{IEEEproof}

As mentioned earlier, this is the first time that the optimal value of $\rho$ minimizing the power consumption is given in explicit form for a generic heuristic precoding matrix $\mathbf{V}$ defined as in \eqref{22.101}. Most of the existing works have only looked for the value of $\rho$ that maximizes the sum rate of the network (see for example \cite{Wagner12}).

From the results of Theorem \ref{theorem_general_V}, the optimal value of $\rho$ for RZF or PA-RZF easily follows.

\begin{corollary}\label{corollary_RZF}
If a RZF precoder is used and $N,K \rightarrow \infty$ with $c \in (0,1]$, then
\begin{align}\label{29.1}
\rho_{\rm{RZF}}^\star = \frac{1}{\mu^\star} - \frac{1}{N}\sum\limits_{i=1}^K\frac{l(\mathbf{x}_{i})}{1+l(\mathbf{x}_{i}) \mu^\star}
\end{align}
with $\mu^\star$ being solution of the following fixed point equation:
\begin{align}\label{30.1}
\!\!\!\mu^\star \!\!= \!\!\left(\sum\limits_{i=1}^K \frac{l(\mathbf{x}_{i}) \gamma_i}{\left(1 + l(\mathbf{x}_{i})\mu^\star\right)^3}\right)\!\!\!\left(\sum\limits_{i=1}^K \frac{\left(l(\mathbf{x}_{i})\right)^2}{\left(1 + l(\mathbf{x}_{i})\mu^\star\right)^3}\right)^{-1}\!\!\!.
\end{align}
\end{corollary}

\begin{IEEEproof}
The proof easily follows from the results of Theorem 1 setting $\alpha_i =1$ for $i=1,2,\ldots,K$.
\end{IEEEproof}

\begin{corollary}
If a PA-RZF precoder is used and $N,K \rightarrow \infty$ with $c \in (0,1]$, then
\begin{align}\label{32.1}
\rho_{\rm{PA-RZF}}^\star = \frac{1}{\beta} - \frac{c}{1 +\beta}
\end{align}
with $\beta >0$ being the average target SINR given by
\begin{align}\label{34}
\beta = \frac{1}{K}\sum\limits_{k=1}^K\gamma_k.
\end{align}
The deterministic equivalent of the minimum transmit power reduces to
\begin{align}\label{36.10}
\overline P_{\rm{PA-RZF}}= {cA\sigma^2}\left(1 - c \frac{\beta}{1+\beta}\right)^{-1}.
\end{align}
\end{corollary}

\begin{IEEEproof}
The result follows directly from Theorem 1 setting $\alpha_i =1/l(\mathbf{x}_{i})$ for $i=1,2,\ldots,K$.
\end{IEEEproof}

\vspace{0.1cm}
Interestingly, the above results can be used to prove under which conditions RZF and PA-RZF are optimal.
\begin{corollary}
If condition \eqref{zeta} holds true, then RZF becomes the optimal linear precoder in the asymptotic regime.
\end{corollary}
\begin{IEEEproof}
From \eqref{zeta}, it follows that $l(\mathbf{x}_{k})\gamma_k = (l(\mathbf{x}_{k}))^2 \zeta$. Plugging this result into \eqref{30.1} yields $\mu^\star = \zeta$ from which we get
\begin{align}
\rho_{\rm{RZF}}^\star = \frac{1}{\zeta}\left(1 - \frac{1}{N}\sum\limits_{i=1}^K\frac{\gamma_i}{1+\gamma_i}\right)
\end{align}
by simple manipulations. Plugging this result into \eqref{28} we obtain
\begin{align}\label{39.AA}
\mathbf{V}_{\rm{RZF}}^\star=\frac{1}{\rho_{\rm{RZF}}^\star}\left(\frac{1}{\rho_{\rm{RZF}}^\star}\sum\limits_{i=1}^K\mathbf{  w}_i\mathbf{  w}_i^H + N\mathbf{I}_N\right)^{-1}\mathbf{  H}\sqrt{\mathbf{P}}
\end{align}
which is equal to \eqref{22} after replacing $\lambda_k^\star$ with $\overline \lambda_k$ in \eqref{26.2.D}.
\end{IEEEproof}

\begin{corollary}If condition \eqref{gamma} holds true, then PA-RZF becomes the optimal linear precoder in the asymptotic regime
\end{corollary}
\begin{IEEEproof}If $\boldsymbol{\gamma} = \gamma \boldsymbol{1}_k$, then \eqref{32.1} reduces to $\rho_{\rm{PA-RZF}}^\star = \frac{1}{\gamma} - \frac{c}{1 + \gamma}$
and $\mathbf{V}_{\rm{PA-RZF}}^\star$ in \eqref{31} becomes equivalent to \eqref{22} after replacing $\lambda_k^\star$ with $\overline\lambda_k$ given by \eqref{26.2}.
\end{IEEEproof}



\begin{figure}
  \begin{center}
    \psfrag{xlabel}[c][b]{\footnotesize{Rate per user $r$ [bit/s/Hz]}}
    \psfrag{ylabel}[c][t]{\footnotesize{Average transmit power [Watt]}}
    \psfrag{data1}[l][c]{\!\!\!\scriptsize{ZF}}
    \psfrag{data2}[l][c]{\!\!\!\scriptsize{RZF}}
    \psfrag{data3}[l][c]{\!\!\!\scriptsize{PA-RZF}}
    \psfrag{data4}[l][c]{\!\!\!\scriptsize{A-OLP}}
        \psfrag{data5}[l][c]{\!\!\!\scriptsize{OLP}}
    \includegraphics[width=0.95\columnwidth]{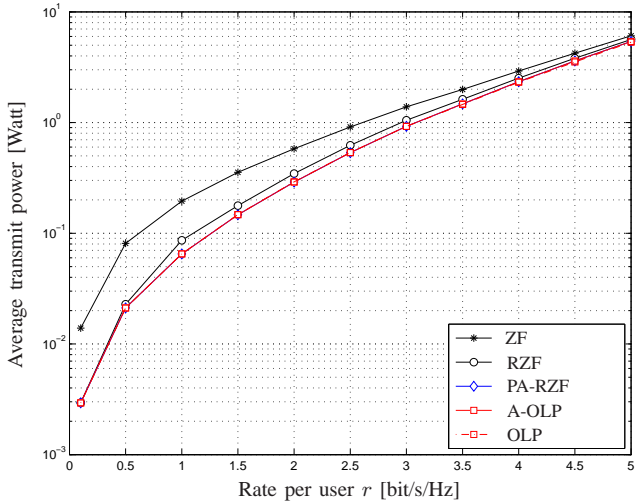}
  \caption{Average transmit power in Watt vs. the rate per user $r$ in bit/s/Hz when $K =8$ and $N=10$.}
  \label{fig1}
  \end{center} \vspace{-0.7cm}
\end{figure}

\section{Numerical Results}

In this section, Monte Carlo simulations are used to validate the analysis in the asymptotic regime and to make comparisons between optimal linear precoding and different heuristic precoding techniques. We assume that the UEs are uniformly distributed in a circular cell with radius $D = 250$ m and minimum distance $D_{\min} = 15$ m. Moreover, we consider a system in which the large-scale fading is dominated by the path-loss \cite{LTE2010b}. This is modelled as $l(\vect{x}) = {d_0}/{\| \vect{x} \|^{\kappa}}$ for $\|\vect{x}\|\geq D_{\min}$
where $\kappa \geq 2$ is the path-loss exponent and the constant $d_0>0$ regulates the channel attenuation at distance $D_{\min}$. In all subsequent simulations, we set $\kappa = 3.76$ and $d_0 = 10^{-3.53}$. In addition, the transmission bandwidth is $W = 10$ MHz and the total noise power $W\sigma^2$ is $-104$ dBm.

We begin by considering a cellular network in which the same rate $r$ in bit/s/Hz must be guaranteed to each UE. This amounts to saying that $\gamma_k = \gamma = 2^{r} -1$ for $k=1,2,\ldots,K$.
Fig. \ref{fig1} illustrates the average transmit power in Watt with $K=8$ and $N=10$ when $r$ spans the interval from $0.1$ to $5$ bit/s/Hz. The curves labelled OLP and A-OLP refer to the performance of the optimal and asymptotically optimal linear precoders, respectively. On the other hand, ZF refers to the classical zero-forcing precoder. From the results of Fig. \ref{fig1}, it follows that OLP and A-OLP have substantially the same performance. As pointed out in Remark 3, PA-RZF provides the same performance of A-OLP. While PA-RZF achieves only a marginal gain compared to RZF, a substantial power reduction is obtained with respect to ZF for moderate values of $r$. The mean-square-error of the effective user rates (not reported here for space limitations) is found to be smaller than $2\%$ meaning that the performance loss is reasonably negligible.

\begin{figure}
  \begin{center}
    \psfrag{xlabel}[c][b]{\footnotesize{$N$}}
    \psfrag{ylabel}[c][t]{\footnotesize{Average transmit power [Watt]}}
    \psfrag{data1}[l][c]{\!\!\!\scriptsize{ZF}}
    \psfrag{data2}[l][c]{\!\!\!\scriptsize{RZF}}
    \psfrag{data3}[l][c]{\!\!\!\scriptsize{PA-RZF}}
    \psfrag{data4}[l][c]{\!\!\!\scriptsize{A-OLP}}
        \psfrag{data5}[l][c]{\!\!\!\scriptsize{OLP}}
    \includegraphics[width=0.95\columnwidth]{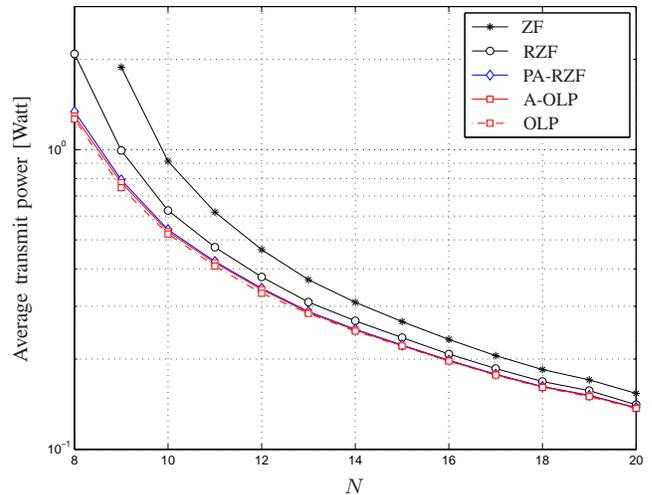}
  \caption{Average transmit power in Watt vs. $N$ when $K = 8$ and the user rates takes values within the interval $[2, 3]$ bit/s/Hz.}
  \label{fig2}
  \end{center}\vspace{-0.7cm}
\end{figure}

Fig. \ref{fig2} plots the average transmit power in Watt vs. $N$ when $K =8$ and the user rates $\{r_k\}$ are randomly taken within the interval $[2, 3]$ bit/s/Hz. Although different rates are requested by the UEs, PA-RZF has substantially the same performance of A-OLP for any value of $N$. A significant gap is observed with respect to ZF for values of $N$ in the order of $K$, while all the schemes guarantee basically the same performance when $N$ becomes larger.

\vspace{-.2cm}
\section{Conclusions}

In this work, we have focused on a single-cell multi-user MIMO system and have studied the problem of designing linear precoding techniques for minimizing the total power consumption while satisfying a set of target SINRs. The solution to this problem is generally given by solving a set of fixed-point equations, which is cumbersome in large-scale MIMO systems. 
To simplify the analysis and overcome complexity issues, we have resorted to the asymptotic regime in which the number of antennas and users grow large with a given ratio. The asymptotic solutions to the fixed-point equations have been given in closed form, thereby providing insights on the optimal precoding structure. In particular, we have used these results to prove that the conventional RZF precoding technique is the optimal one in the asymptotic regime when the ratio between the SINR requirement and the average channel attenuation is the same for all UEs. A position-aware RZF (PA-RZF) precoding that exploits statistical knowledge of the UE positions has been shown to be asymptotically optimal in realistic scenarios where the SINR constraints are the same but the path-losses are different.

\vspace{-0.1cm}
\section*{Appendix}
If $\mathbf{V}$ takes the generic heuristic form in \eqref{22.101}, then for any given $\boldsymbol{\alpha}= [\alpha_1,\alpha_2,\ldots,\alpha_K]^T$ and $\rho$ the following lemma can be proved using the results of Theorem 1 in \cite{Wagner12}.

\begin{lemma}
If $N,K \rightarrow \infty$ with $c \in (0,1]$, then
\begin{align}\label{19}
 P  - \overline P &\mathop {\longrightarrow}\limits^{a.s.} 0
\\\label{20}
{\rm{SINR}}_k  - {\rm{\overline {SINR}}}_k &\mathop {\longrightarrow}\limits^{a.s.} 0
\end{align}
where $\overline P$ and $ {\rm{\overline {SINR}}}_k$ are given by
\begin{align}\label{A16}
\overline P= \frac{c \mu^\prime}{K}\sum\limits_{i=1}^K\frac{p_il(\mathbf{x}_{i})}{\left(1 + \alpha_il(\mathbf{x}_{i})\mu\right)^2}
\end{align}
\begin{equation}\label{A17}
{\rm{\overline {SINR}}}_k= \frac{p_kl(\mathbf{x}_{k})\mu^2}{\overline P + \frac{\sigma^2}{l(\mathbf{x}_{k})} \left[1+\alpha_k l(\mathbf{x}_{k})\mu\right]^2}
\end{equation}
where $\mu$ is the solution of the following fixed point equation
\begin{align}\label{mu}
\mu = \left(\frac{1}{N}\sum\limits_{i=1}^K\frac{ \alpha_il(\mathbf{x}_{i})}{1+\alpha_i l(\mathbf{x}_{i}) \mu} + \rho\right)^{-1}
\end{align}
and $\mu^\prime$ in \eqref{A16} is its derivative with respect to $\rho$.
\end{lemma}

To proceed further, we set ${\rm{\overline {SINR}}}_k= \gamma_k$ for $k=1,2,\ldots,K$ and compute the corresponding power $\overline P$. 

\begin{lemma}
If ${\rm{\overline {SINR}}}_k$ is set equal to $\gamma_k$ for $k=1,2,\ldots,K$, then $\overline P$ is found to be
\begin{align}\label{12}
\overline P  =  \frac{ c  A\sigma^2}{1 - \mu^2F_2-c B}
\end{align}
with $A$ and $\mu$ being given by \eqref{26} and \eqref{mu} whereas $B$ and $F_2$ takes the form in \eqref{36.1001} and \eqref{36.1002}. 
\end{lemma}

%
%
%
%

\begin{IEEEproof}
Setting ${\rm{\overline {SINR}}}_k$ in \eqref{A17} equal to $\gamma_k$ leads to
\begin{align}\label{A19}
p_k = \frac{\gamma_k}{f_k(\rho)} = \frac{{\gamma_k\left\{\overline P + \frac{\sigma^2}{l(\mathbf{x}_{k})} \left[1+\alpha_kl(\mathbf{x}_{k})\mu\right]^2\right\}}}{l(\mathbf{x}_{k})\mu^2}
\end{align}
which used in \eqref{A16} yields
\begin{align}
\overline P= -\frac{c\mu^\prime}{\mu^2} \frac{1}{K}\sum\limits_{i=1}^K \frac{{ \gamma_i\left\{\overline P + \frac{\sigma^2}{l(\mathbf{x}_{i})} \left[1+\alpha_il(\mathbf{x}_{i})\mu\right]^2\right\}}}{\left(1 + \alpha_il(\mathbf{x}_{i})\mu\right)^2}.
\end{align}
Solving with respect to $\overline P$ produces
\begin{align}\label{A20}
\overline P =-\frac{c\mu^\prime }{\mu^2} \frac{ A\sigma^2}{1+c \frac{\mu^\prime }{\mu^2}B}.
\end{align}
Observing that the derivative of $\mu$ in \eqref{mu} with respect to $\rho$ is $\mu^\prime = \frac{\mu^2}{1 - \mu^2F_2}$
the result in \eqref{12} follows from \eqref{A20}.
\end{IEEEproof}

Taking the derivative of $\overline P$ in \eqref{12} with respect to $\rho$ yields (the mathematical details are omitted for space limitations)
\begin{align}\label{A23}
\overline P^\prime = 2c^2 A\sigma^2 \mu^\prime\frac{\mu \mathcal A - \mathcal B }{\left(1 - \mu^2F_2-c B\right)^2}
\end{align}
with
\begin{align}\nonumber
\mathcal A = \frac{1}{K}\sum\limits_{i=1}^K \frac{\left(\alpha_il(\mathbf{x}_{i})\right)^2}{\left(1 + \alpha_il(\mathbf{x}_{i})\mu\right)^3} \quad 
\mathcal B = \frac{1}{K}\sum\limits_{i=1}^K \frac{\alpha_il(\mathbf{x}_{i})\gamma_i}{\left(1 + \alpha_il(\mathbf{x}_{i})\mu\right)^3}.
\end{align}
From \eqref{A23}, it turns out that the optimal $\mu$ is such that $\mu^\star = {\mathcal B}/{\mathcal A}$
from which using \eqref{mu} the optimal $\rho$ is found to be in the form of \eqref{23.1} in the text.
%

\vspace{-0.1cm}
\section*{Acknowledgment}
The authors thank Dr.~Romain Couillet for helpful discussions on the large system analysis of the optimal linear precoding and in particular for the results of Theorem 1.

\vspace{-0.1cm}
\bibliographystyle{IEEEtran}
\bibliography{IEEEabrv,refs}

\begin{thebibliography}{10}
\providecommand{\url}[1]{#1}
\csname url@samestyle\endcsname
\providecommand{\newblock}{\relax}
\providecommand{\bibinfo}[2]{#2}
\providecommand{\BIBentrySTDinterwordspacing}{\spaceskip=0pt\relax}
\providecommand{\BIBentryALTinterwordstretchfactor}{4}
\providecommand{\BIBentryALTinterwordspacing}{\spaceskip=\fontdimen2\font plus
\BIBentryALTinterwordstretchfactor\fontdimen3\font minus
  \fontdimen4\font\relax}
\providecommand{\BIBforeignlanguage}[2]{{%
\expandafter\ifx\csname l@#1\endcsname\relax
\typeout{** WARNING: IEEEtran.bst: No hyphenation pattern has been}%
\typeout{** loaded for the language `#1'. Using the pattern for}%
\typeout{** the default language instead.}%
\else
\language=\csname l@#1\endcsname
\fi
#2}}
\providecommand{\BIBdecl}{\relax}
\BIBdecl

\bibitem{Qinghua2010}
Q.~Li, G.~Li, W.~Lee, M.~il~Lee, D.~Mazzarese, B.~Clerckx, and Z.~Li, ``{MIMO}
  techniques in {WiMAX} and {LTE}: a feature overview,'' \emph{IEEE Commun.
  Mag.}, vol.~48, no.~5, pp. 86--92, May 2010.

\bibitem{Weingarten2006}
H.~Weingarten, Y.~Steinberg, and S.~Shamai, ``The capacity region of the
  gaussian multiple-input multiple-output broadcast channel,'' \emph{IEEE
  Trans. Inf. Theory}, vol.~52, no.~9, pp. 3936--3964, Sept 2006.

\bibitem{Gershman2010}
A.~Gershman, N.~Sidiropoulos, S.~Shahbazpanahi, M.~Bengtsson, and B.~Ottersten,
  ``Convex optimization-based beamforming,'' \emph{IEEE Signal Process. Mag.},
  vol.~27, no.~3, pp. 62--75, May 2010.

\bibitem{Bjornson:2014mag}
E.~Bj{\"{o}}rnson, M.~Bengtsson, and B.~Ottersten, ``Optimal multiuser transmit
  beamforming: A difficult problem with a simple solution structure [lecture
  notes],'' \emph{IEEE Signal Processing Magazine}, vol.~31, no.~4, pp.
  142--148, July 2014.

\bibitem{Schubert2004a}
M.~Schubert and H.~Boche, ``Solution of the multiuser downlink beamforming
  problem with individual {SINR} constraints,'' \emph{IEEE Trans. Veh. Tech.},
  vol.~53, no.~1, pp. 18--28, Jan. 2004.

\bibitem{Wiesel06}
A.~Wiesel, Y.~Eldar, and S.~Shamai, ``Linear precoding via conic optimization
  for fixed {MIMO} receivers,'' \emph{IEEE Trans. Signal Process.}, vol.~54,
  no.~1, pp. 161--176, Jan. 2006.

\bibitem{Yu:2007ja}
W.~Yu and T.~Lan, ``Transmitter optimization for the multi-antenna downlink
  with per-antenna power constraints,'' \emph{IEEE Trans. Signal Process.},
  vol.~55, no.~6, pp. 2646--2660, Jun. 2007.

\bibitem{Bjornson2012a}
E.~Bj{\"{o}}rnson, G.~Zheng, M.~Bengtsson, and B.~Ottersten, ``Robust monotonic
  optimization framework for multicell {MISO} systems,'' \emph{IEEE Trans.
  Signal Process.}, vol.~60, no.~5, pp. 2508--2523, May 2012.

\bibitem{Chen2011}
Y.~Chen, S.~Zhang, S.~Xu, and G.~Li, ``Fundamental trade-offs on green wireless
  networks,'' \emph{IEEE Commun. Mag.}, vol.~49, no.~6, pp. 30--37, June 2011.

\bibitem{Bengtsson2001a}
M.~Bengtsson and B.~Ottersten, ``Optimal and suboptimal transmit beamforming,''
  in \emph{Handbook of Antennas in Wireless Communications}, L.~C. Godara,
  Ed.\hskip 1em plus 0.5em minus 0.4em\relax CRC Press, 2001.

\bibitem{Lakshminaryana2010}
S.~Lakshminaryana, J.~Hoydis, M.~Debbah, and M.~Assaad, ``Asymptotic analysis
  of distributed multi-cell beamforming,'' in \emph{IEEE 21st International
  Symposium on Personal Indoor and Mobile Radio Communications (PIMRC)}, Sept
  2010, pp. 2105--2110.

\bibitem{Huang2012}
Y.~Huang, C.~W. Tan, and B.~Rao, ``Large system analysis of power minimization
  in multiuser {MISO} downlink with transmit-side channel correlation,'' in
  \emph{International Symposium on Information Theory and its Applications
  (ISITA)}, Oct. 2012, pp. 240--244.

\bibitem{Zakhour2012}
R.~Zakhour and S.~Hanly, ``Base station cooperation on the downlink: Large
  system analysis,'' \emph{IEEE Trans. Inf. Theory}, vol.~58, no.~4, pp.
  2079--2106, Apr. 2012.

\bibitem{Asgharimoghaddam14}
\BIBentryALTinterwordspacing
A.~T. H.~Asgharimoghaddam and N.~Rajatheva, ``Decentralizing the optimal
  multi-cell beamforming via large system analysis,'' in \emph{Proceedings of
  the IEEE International Conference on Communications}, Sydney, Australia, June
  2014. [Online]. Available: \url{http://arxiv.org/abs/1310.3843}
\BIBentrySTDinterwordspacing

\bibitem{Romain2014}
R.~Couillet and M.~McKay, ``Large dimensional analysis and optimization of
  robust shrinkage covariance matrix estimators,'' \emph{Journal of
  Multivariate Analysis}, vol. 131, no.~0, pp. 99 -- 120, 2014.

\bibitem{Wagner12}
S.~Wagner, R.~Couillet, M.~Debbah, and D.~T.~M. Slock, ``Large system analysis
  of linear precoding in correlated {MISO} broadcast channels under limited
  feedback,'' \emph{IEEE Transactions on Information Theory}, vol.~58, no.~7,
  pp. 4509--4537, July 2012.

\bibitem{Evans2013}
R.~Muharar, R.~Zakhour, and J.~Evans, ``Optimal power allocation and user
  loading for multiuser {MISO} channels with regularized channel inversion,''
  \emph{IEEE Trans. Commun.}, vol.~61, no.~12, pp. 5030--5041, Dec. 2013.

\bibitem{Bjornson:2013kc}
E.~Bj{\"{o}}rnson and E.~Jorswieck, ``Optimal resource allocation in
  coordinated multi-cell systems,'' \emph{Foundations and Trends in
  Communications and Information Theory}, vol.~9, no. 2-3, pp. 113--381, 2013.

\bibitem{Sanguinetti_JSAC_14}
\BIBentryALTinterwordspacing
L.~Sanguinetti, A.~L. Moustakas, and M.~Debbah, ``Interference management in
  {5G} reverse {TDD HetNets}: A large system analysis,'' \emph{submitted to
  IEEE J. Sel. Areas Commun.}, July 2014. [Online]. Available:
  \url{http://arxiv.org/abs/1407.6481}
\BIBentrySTDinterwordspacing

\bibitem{Sanguinetti_2014_Mobility}
\BIBentryALTinterwordspacing
L.~Sanguinetti, A.~L. Moustakas, E.~Bj{\"o}rnson, and M.~Debbah, ``Large system
  analysis of the energy consumption distribution in multi-user {MIMO} systems
  with mobility,'' \emph{submitted to IEEE Trans. Wireless Commun.}, June 2014.
  [Online]. Available: \url{http://arxiv.org/abs/1406.5988}
\BIBentrySTDinterwordspacing

\bibitem{Rusek2013a}
F.~Rusek, D.~Persson, B.~Lau, E.~Larsson, T.~Marzetta, O.~Edfors, and
  F.~Tufvesson, ``Scaling up {MIMO}: Opportunities and challenges with very
  large arrays,'' \emph{IEEE Signal Process. Mag.}, vol.~30, no.~1, pp. 40--60,
  Jan. 2013.

\bibitem{LTE2010b}
\emph{Further advancements for {E-UTRA} physical layer aspects (Release
  9)}.\hskip 1em plus 0.5em minus 0.4em\relax {3GPP} {TS} 36.814, Mar. 2010.

\end{thebibliography}
\end{document}